\documentstyle[preprint,aps]{revtex}
\begin{document}
\title{Superintegrable Systems, Multi-Hamiltonian Structures and Nambu Mechanics in
an Arbitrary Dimension }
\author{A. Te\u{g}men \thanks{%
Electronic mail: tegmen@science.ankara.edu.tr} and A. Ver\c{c}in \thanks{%
Electronic mail: vercin@science.ankara.edu.tr}}
\address{Department of Physics, Ankara University, Faculty of Sciences,\\
06100, Tando\u{g}an-Ankara, Turkey.}
\maketitle

\begin{abstract}
A general algebraic condition for the functional independence of $2n-1$
constants of motion of an $n$-dimensional maximal superintegrable
Hamiltonian system has been proved for an arbitrary finite $n$. This makes
it possible to construct, in a well-defined generic way, a normalized Nambu
bracket which produces the correct Hamiltonian time evolution. Existence and
explicit forms of pairwise compatible multi-Hamiltonian structures for any
maximal superintegrable system have been established. The Calogero-Moser
system, motion of a charged particle in a uniform perpendicular magnetic
field and Smorodinsky-Winternitz potentials are considered as illustrative
applications and their symmetry algebras as well as their Nambu formulations
and alternative Poisson structures are presented.
\end{abstract}

{\bf PACS}: 02.30.Ik, 02.40.Yy, 45.20.Jj

\tightenlines
\newpage

\section{Introduction}

Nambu mechanics is a generalization of the Hamiltonian formulation of
classical mechanics in that it replaces the usual binary Poisson bracket
(PB) to higher order $n$-ary bracket, generically called Nambu bracket (NB),
and specifies the dynamics in terms of $n-1$ ``generalized Hamiltonian''
functions \cite{Nambu,Takhtajan}. The original motivation of Nambu was to
show that the Hamiltonian mechanics is not the only formulation that makes a
statistical mechanics possible. Relevance of Nambu mechanics to membrane
theory has been put forward and a form of quantized Nambu mechanics has been
purposed as a nonlinear generalization of geometric formulation of quantum
mechanics \cite{Minic,Hughston}. Nambu formulation may also give some
insights into the theory of higher order algebraic structures and their
possible physical significance \cite{Marmo}. Unfortunately, up to now only
few examples of dynamical systems which admit Nambu formulation have been
given. The Euler equations for three dimensional $(3D)$ rigid body were the
only example given by Nambu. Then, the equations of Nahm system in the
theory of static $SU(2)$ monopoles were realized in this formulation \cite
{Takhtajan,Chak}. Connection between Nambu mechanics and so-called
superintegrable systems has been the subject of some recent studies \cite
{Chatterjee,Hietarinta1,Gonera,Zachos}.

A Hamiltonian system of $n$ degrees of freedom is called to be completely
integrable, in the Liouville-Arnold sense, if it admits $n$ functionally
independent, globally defined constants of motion in involution (i.e.,
commuting with respect to PB) \cite{Arnold,Hietarinta2}. A completely
integrable system is called superintegrable if it allows $k$ additional
constants of motion. Not all constants of motion of a superintegrable system
can be in involution but they must be functionally independent, otherwise
the extra invariants are trivial. Superintegrability is said to be minimal
if $k=1$, and maximal if $k=n-1$ \cite{Evans,Evans2,Wol}. An $nD$ maximally
superintegrable Hamiltonian system can be specified by the following set 
\begin{equation}
SI_{H}(2n-1)=\{H,\;H_{i},\;A_{j}:\; \{H,H_{i}\}=0= \{H_{i},H_{j}\},\;
\{H,A_{i}\}=0,\;\Gamma \neq 0\}\;,
\end{equation}
where $H$ is the Hamiltonian of the system, $H_{i},
\;A_{j};\;i,\;j=1,2,\dots, n-1$ are the additional constants of motion and $%
\Gamma $ denotes the following $(2n-1)$-form 
\begin{equation}
\Gamma =dH\wedge dH_{1}\wedge \cdots \wedge dH_{n-1}\wedge dA_{1}\wedge
\cdots \wedge dA_{n-1}.
\end{equation}
Here $d$ and $\wedge $ denote the usual exterior derivative and exterior
product of Cartan calculus. For functional independence $\Gamma $ must be
different from zero on a dense subset of the underlying symplectic manifold
endowed with the PB $\{,\}$.

In this paper three main points concerning the fundamental structure of a $%
SI_{H}(2n-1)$ system for any finite $n$ are established. We shall first
prove that the constants of motion of a $SI_{H}(2n-1)$ system are
functionally independent where $(n-1)\times (n-1)$ matrix $B$ with elements $%
B_{ij}=\{H_{i},A_{j}\}$ is nonsingular. Secondly we shall construct the
normalized NB which produces the correct Hamiltonian time evolutions. This
means that all $nD$ maximally superintegrable Hamiltonian systems admit
Nambu formulation. We then show that every $SI_{H}(2n-1)$ system admits $%
2n-1 $ multi-Hamiltonian structures. Statements that we have proved make
these facts possible, in a well defined generic way and independently from
the forms of Hamiltonians. Multi-Hamiltonian structures of maximally
superintegrable systems were considered for the first time in Ref.\cite
{Gonera} by using a different approach from ours. In this context, our
geometric proofs of Jacobi identity and of compatibility condition for
alternative Poisson structures are different but validate and, in a sense,
are complementary to that of Ref.\cite{Gonera}.

In the next section main points of the Nambu mechanics, mainly those needed
for the subsequent investigation are briefly reviewed ( for more details we
refer to \cite{Takhtajan}). In Section III a coordinate-free form of
canonical NB and normalized NB are introduced, and the Jacobi identity for a
normalized binary bracket induced from NB is established. The general
algebraic condition for the functional independence is proved in section IV.
Nambu formulation of a $SI_{H}(2n-1)$ system is established in section V
where we also point out some general facts concerning the structure of
symmetry algebras of maximally superintegrable systems. Multi-Hamiltonian
structures are taken up in section VI. As applications the Calogero-Moser
system, motion of a charged particle in a uniform perpendicular magnetic
field (this will be referred to as (classical) Landau problem) and
Smorodinsky-Winternitz potentials are considered in the final section where
their symmetry algebras and explicit forms of their alternative Hamiltonian
structures are established.

Our further notational conventions are as follows. We shall denote the
linear spaces of all vector fields and (differential) $p$-forms on a smooth
manifold $M$ of dimension $n$, respectively, by ${\cal X}(M)$ and $%
\Lambda^{p}(M),\;0\leq p \leq n $. Together with their commutative (and
associative) algebra structure with respect to usual point-wise product, the
linear space of all smooth functions ($0$-forms) defined on $M$ will be
represented by ${\cal A}$. We shall denote the vector fields by bold face
letters, adopt the Einstein summation convention over repeated pair of
contravariant and covariant indices and use the shorthand ${\cal A}^{\otimes
^{n}}={\cal A}\otimes\cdots \otimes {\cal A}$ ($n$ times).

\section{Nambu Mechanics}

NB of order $n$ is the real multilinear map $\{,\;...,\;\}:\;{\cal A}%
^{\otimes^{n}}\rightarrow {\cal A}$ which has, for all $f_{j},g_{j}\in {\cal %
A}$, the following properties.\newline
\vspace{0,1cm} \newline
{\bf i}. Skew-symmetry 
\begin{equation}
\{f_{1},\;\dots,\;f_{n}\}=(-1)^{\varepsilon }\{f_{\sigma (1)},\;\dots
,\;f_{\sigma (n)}\},
\end{equation}
where $\sigma $ is a member of the permutation group $S_{n}$ and $%
\varepsilon $ is the parity of permutation $\sigma $ ($\varepsilon =1$ for
odd permutations, and $\varepsilon =0$ for even permutations). \newline
{\bf ii}. Derivation (the Leibniz rule) 
\begin{equation}
\{f_{1}f_{2},f_{3},\;\dots,\;f_{n+1}\}=f_{1}\{f_{2},f_{3},\;\dots
,\;f_{n+1}\}+f_{2}\{f_{1},f_{3},\;\dots,\;f_{n+1}\}.
\end{equation}
\newline
{\bf iii}. Fundamental identity (a kind of generalized Jacobi identity) 
\begin{equation}
\{f_{1},\;\dots,\;f_{n-1},\{g_{1},\;\dots,\;g_{n}\}\}
=\sum_{k=1}^{n}\{g_{1},\;\dots,\;\{f_{1},\;\dots
,\;f_{n-1},g_{k}\},\;\dots,\;g_{n}\}.
\end{equation}
With respect to NB, ${\cal A}$ acquires another algebra structure henceforth
denoted by ${\cal A}_{N}$.

Nambu dynamics is determined by $n-1$ Hamiltonian functions $%
h_{1},\;\dots,\;h_{n-1}\in {\cal A}_{N}$ and is described, for any $f \in 
{\cal A}_{N} $, by the Nambu-Hamilton (NH) equations of motion 
\begin{equation}
\frac{df}{dt}=X_{NH}(f)=\{f, h_{1},\;\dots,\;h_{n-1} \},
\end{equation}
where $X_{NH}$ is called the NH vector field corresponding to $%
h_{1},\;\dots,\;h_{n-1}$.

NB of order $n$ induces infinite family of lower order NB, including the
family of Poisson structures, all of which satisfy corresponding fundamental
identities (FIs) that follow from (5). Below we shall concentrate only on
the induced Poisson structures. For a fixed set of $n-2$ Hamiltonian
functions $f_{i}\in {\cal A}_{N}$ we define the Nambu induced PB as follows 
\begin{equation}
\{f,g\}_{NP}=\{f,g,f_{1},\;\dots ,\;f_{n-2}\},
\end{equation}
where $f,\;g\in {\cal A}_{N}$ are arbitrary. If in Eq. (5) we take 
\[
f_{n-1}=f,:g_{n-1}=g,\;g_{n}=h,\;g_{i}=f_{i};\;i=1,\;\dots ,\;n-2\; 
\]
then, by virtue of (3), the first $n-2$ terms at the right hand side (5)
vanish and we get 
\begin{equation}
\{f,\{g,h\}_{NP}\}_{NP}+cp=0,
\end{equation}
where $cp$ stands for cyclic permutations. Eq. (8) reveals the fact that $%
\{,\}_{NP}$ satisfies the Jacobi identity. Note that all the fixed functions
in the definition (7) are Casimirs of $\{,\}_{NP}$, that is, $%
\{f_{i},f\}_{NP}=0$ for all $f_{i},\;i=1,\;\dots ,\;n-2$ and $f\in {\cal A}%
_{N}$.

\section{Canonical NB and Normalized NB}

The problem of constructing concrete realizations of NB is of great
importance. In the case of $M={\bf R}^{n}$ the following form, called the
canonical NB 
\begin{equation}
\{f_{1},\;\dots,\;f_{n}\}=\frac{\partial (f_{1},\;\dots,\;f_{n})}{\partial
(x^{1},\;\dots,\;x^{n})}\;,
\end{equation}
was provided by Y. Nambu. Here $(x^{1},\;\dots,\;x^{n})$ denote the local
coordinates of ${\bf R}^{n}$ and the right hand side stands for the Jacobian
of the mapping $f=(f_{1},\;\dots,\;f_{n}):{\bf R}^{n}\rightarrow {\bf R}^{n}$%
.

We shall now introduce a coordinate-free expression of the canonical NB that
provides a considerable ease in proving the technical points of the paper.
For this purpose we first associate the $n-1$-form 
\begin{equation}
\gamma =dh_{1}\wedge \cdots \wedge dh_{n-1}\; ,
\end{equation}
to $n-1$ Hamiltonian functions $h_{j}\in {\cal A}_{N}$. We then recall the
Hodge map $^{\star }:\Lambda ^{p}({\bf R}^{n})\rightarrow \Lambda ^{n-p}(%
{\bf R}^{n})$ defined for any p-form $w=(1/p!)w_{i_{1}\;\dots
\;i_{p}}dx^{i_{1}}\wedge \cdots\wedge dx^{i_{p}}$ as follows \cite{Thirring} 
\begin{eqnarray}
^{\star }w =\frac{1}{p!(n-p)!}{{\epsilon }^{i_{1}\;\cdots \;i_{p}}}%
_{i_{p+1}\;\dots \;i_{n}}w_{i_{1}\;\dots\;i_{p}}dx^{i_{p+1}}\wedge \cdots
\wedge dx^{i_{n}},  \nonumber
\end{eqnarray}
where $w_{i_{1}\;\dots\;i_{p}}$ are antisymmetric cotravariant components of 
$w$ and $\epsilon ^{i_{1}\;\dots\;i_{n}}$ with $\epsilon^{1\;\dots\;n}=1$ is
the $nD$ completely antisymmetric Levi-Civita symbol. Note that with respect
to ${\cal A}$ the Hodge map is linear and exterior product is bilinear. It
is now obvious that, in local coordinates 
\begin{equation}
^{\star}(df\wedge\gamma)=\{f,h_{1},\;\dots,\; h_{n-1}\}= \frac{%
\partial(f,\;h_{1},\;\dots,\;h_{n-1})}{\partial(x^{1},\;\dots,\;x^{n})}.
\end{equation}
In that case, multilinearity, antisymmetry and derivation properties of the
canonical NB are direct results of the linearity of the Hodge map with
respect to ${\cal A}$, and of the well-known properties of $\wedge$-product
and $d$. We can also associate the $n-1$-form $\beta=\beta_{i_{1}\;\dots\;
i_{n-1}} dx^{i_{1}}\wedge \cdots \wedge dx^{i_{n-1}}/(n-1)!$ to the vector
field $\bbox{\beta}=(\beta^{1},\;\dots,\;\beta^{n})$ with components $%
\beta^{k}=\epsilon^{ki_{1}\;\dots\;i_{n-1}} \beta_{i_{1}\;\dots
\;i_{n-1}}/(n-1)!$. Then, (11) can also be written as 
\begin{equation}
^{\star}(df\wedge\gamma)=\bbox{\gamma}\cdot \bbox{\nabla}f,
\end{equation}
where $\bbox{\nabla}$ stands for $n$-dimensional gradient operator and $%
``\cdot"$ denotes the usual inner product of ${\bf R}^{n}$. The fundamental
identity can be verified by taking $f=\{ g_{1},\dots ,g_{n} \}$ in Eq. (12).

We should note that Eq. (12) implies 
\begin{equation}
X_{NH}=(-1)^{n-1}{}^{\star}(\gamma\wedge d)= \bbox{\gamma}\cdot \bbox{\nabla}%
,
\end{equation}
for the NH vector field corresponding to $h_{1},\;\dots,\;h_{n-1}$. As
illustrative examples, let us consider the cases $n=2,\;3$. The vector
fields corresponding to $\gamma=dh$ in the case of $n=2$ and $%
\gamma^{\prime}=dh_{1}\wedge dh_{2}$ in the case of $n=3$ are easily found
to be 
\begin{eqnarray}
\bbox{\gamma}=(\partial_{2}h,\;-\partial_{1}h),\qquad \bbox{\gamma}^{\prime}=%
\bbox{\nabla}h_{1}\bbox{\times}\bbox{\nabla}h_{2} \;,  \nonumber
\end{eqnarray}
where $\partial_{j}=\partial/\partial x_{j}$ and $\bbox{\times}$ denotes the
cross product of ${\bf R}^{3}$. The associated NH vector fields and
canonical NBs can be written as follows 
\begin{eqnarray}
X_{NH}(f)=\partial_{2}h\partial_{1}f-\partial_{1}h\partial_{2}f, \qquad
X_{NH}^{\prime}(f)=(\bbox{\nabla}h_{1}\bbox{\times} \bbox{\nabla}h_{2})\cdot %
\bbox{\nabla}f .  \nonumber
\end{eqnarray}
The first is the usual PB of ${\bf R}^{2}$ and the second is the original NB
first appeared in \cite{Nambu}. In the next two sections we shall generalize
these expressions for a $SI_{H}(2n-1)$ system in the case of arbitrary $n$.

To be precise, from now on we shall adopt, in accordance with related
literature, the following definition : $\;$ If the time evolution equations
of a dynamical system can be written in terms of (canonical) NB then the
system will be called to admit equivalent Nambu formulation. As it will be
apparent in the next two sections in order to get the correct dynamics the
induced PBs must be properly normalized. For this purpose, in terms of $n-2$%
-form 
\begin{equation}
\eta =df_{1}\wedge \;\cdots \;\wedge df_{n-2},
\end{equation}
we define 
\begin{equation}
\{f,g\}_{NP}^{\prime }=C\{f,g,f_{1},\;\dots ,\;f_{n-2}\}=C^{\ast }(df\wedge
dg\wedge \eta ),
\end{equation}
where $C\in {\cal A}$ is, for the time being, an arbitrary function. $C$
will be referred to as the normalization coefficient and will be specified
from the requirement that the NH equation produces the correct time
evolution for any function. The generic form of $C$ will be determined in
the next section, but before that it must be emphasized that $%
\{,\}_{NP}^{\prime }$ satisfies the Jacobi identity for any $C$. To prove
this let us consider 
\begin{equation}
\{h,\{f,g\}_{NP}^{\prime }\}_{NP}^{\prime }=C^{\ast }\{dh\wedge d[C^{\ast
}(df\wedge dg\wedge \eta )]\wedge \eta \}
\end{equation}
and its cyclic permutations for three functions $h,f$ and $g$. Since in a $%
nD $ space we have at most $n$ functional independent functions and since in
(15) we have, apart from $C$, $n+1$ functions, in the most general case one
of them, say $f$, must be functional dependent to others. Hence we can take 
\begin{equation}
df=adh+bdg+\sum_{1}^{n-2}c_{i}f_{i},
\end{equation}
where $a,b$ and $c_{i}$ are arbitrary constants. On substituting this in
(16) and in its $cp$ we immediately see that their sum vanishes.

Obviously, in the case of $n=2$ we have $C=1$. For known $3D$ examples,
namely, for free rigid body and the Nahm equations we also have $C=1$.
However the requirement of nontrivial $C$ is inevitable at least when $n$ is
an even integer greater than two. Although this normalization requirement
have appeared in the literature, its generic form and important implications
were not recognized.

\section{An Algebraic Expression For Functional Independence}

The phase-space of a Hamiltonian system is a $2n$ dimensional symplectic
manifold $M$ on which a symplectic structure is defined by a closed ($%
d\Omega =0$) and nondegenerate symplectic 2-form $\Omega $. Two immediate
implications of nondegeneracy are that \cite{Marsden,Arnold}; $\;$ (i) $M$
is orientable with nowhere vanishing Liouville measure (volume form) 
\begin{equation}
V_{L}=\frac{(-1)^{n(n-1)/2}}{n!}\Omega ^{n},
\end{equation}
where $\Omega ^{n}=\Omega \wedge \cdots \wedge \Omega $ (n times). (ii)
There is a natural isomorphism between the vector fields and 1-forms defined
by $\bbox{\xi}\rightarrow \mu_{\bbox{\xi}}=i_{\bbox{\xi}}\Omega $, where $i_{%
\bbox{\xi}}:\Lambda^{p}(M)\rightarrow \Lambda ^{p-1}(M)$ is called the
interior product operator defined for any p-form $\alpha $ by 
\begin{equation}
(i_{\bbox{\xi}}\alpha )(\bbox{\xi}_{1},\dots ,\bbox{\xi}_{p-1})=\alpha (%
\bbox{\xi},\bbox{\xi}_{1},\dots ,\bbox{\xi}_{p-1}).
\end{equation}
If $\mu _{\bbox{\xi}}$ is exact, that is, if $\mu _{\bbox{\xi}}=df$, then $%
\bbox{\xi}$ is called a Hamiltonian vector field corresponding to $f\in 
{\cal A}$ and henceforth denoted by $\bbox{\xi}_{f} :\; i_{\bbox{\xi}%
_{f}}\Omega =df$. PB on $M$ is defined by $\Omega(\bbox{\xi}_{f},\;\bbox{\xi}%
_{g})=\{f,g\} $. According to Darboux theorem at each point of $M$ there are
local canonical coordinates $(q^{1},\dots ,q^{n},p_{1},\dots ,p_{n})$ in
which $\Omega$ takes the form $\Omega =dq^{j}\wedge dp_{j}$ and leads us to
the following coordinate expressions 
\begin{mathletters}
\begin{eqnarray}
\{ f,g\} &= & \Omega(\bbox{\xi}_{f},\bbox{\xi}_{g})= \partial_{q^{j}}f
\partial_{p_{j}}g -\partial_{p_{j}}f \partial_{q^{j}}g, \\
\bbox{\xi}_{f} &=& \partial_{p_{j}}f \partial_{q^{j}}
-\partial_{q^{j}}f\partial_{p_{j}}, \\
V_{L} &=&dq^{1}\wedge \cdots \wedge dq^{n}\wedge dp_{1}\wedge \cdots \wedge
dp_{n}\;.
\end{eqnarray}

By definition of exterior forms and their $\wedge $-products, the value of $%
\alpha \wedge \beta ;\;\alpha \in \Lambda ^{p},\;\beta \in \Lambda ^{q}$ on $%
p+q$ vectors $\bbox{\xi}_{k}\in {\cal X}(M);k=1,\;2,\;\dots ,\;p+q\leq 2n$,
is given, in the notation of \cite{Arnold}, by 
\end{mathletters}
\begin{equation}
(\alpha \wedge \beta )(\bbox{\xi}_{1},\;\dots ,\;\bbox{\xi}%
_{p+q})=\sum_{S_{p+q}}(-1)^{\varepsilon }\alpha (\bbox{\xi}_{i_{1}},\;\dots
,\;\bbox{\xi}_{i_{p}})\beta (\bbox{\xi}_{j_{1}},\;\dots ,\;\bbox{\xi}%
_{j_{q}}),
\end{equation}
where $i_{1}<\;\cdots \;<i_{p}$ and $j_{1}<\;\cdots \;<j_{q}$ such that $%
(i_{1},\;\dots ,\;i_{p},\;j_{1},\;\dots ,\;j_{q})$ is a permutation of $%
(1,\;2,\;\dots ,\;p+q)$. The summation in Eq. (21) is over all permutations
of the permutation group $S_{p+q}$, provided that the indices are
partitioned into two ordered sets as given above.

By making use of Eq. (21) we now evaluate the Liouville form $V_{L}$ given
by (20c) on $2n$ Hamiltonian vector fields $\bbox{\xi}_{f_{A}};\;A=1,\dots
,\;2n$, as follows 
\begin{eqnarray}
V_{L}(\bbox{\xi}_{f_{1}},\;\dots ,\;\bbox{\xi}_{f_{2n}}) &=&\varepsilon
^{i_{1}\;\cdots \;i_{2n}}dq^{1}(\bbox{\xi}_{f_{i_{1}}})\;\cdots \;dq^{n}(%
\bbox{\xi}_{f_{i_{n}}})dp_{1}(\bbox{\xi}_{f_{i_{n+1}}})\;\cdots \;dp_{n}(%
\bbox{\xi}_{f_{i_{2n}}})  \nonumber \\
&=&(-1)^{n}\varepsilon ^{i_{1}\;\cdots \;i_{2n}}\partial
_{p_{1}}f_{i_{1}}\;\cdots \;\partial _{p_{n}}f_{i_{n}}\partial
_{q^{1}}f_{i_{n+1}}\;\cdots \;\partial _{q^{n}}f_{i_{2n}}  \nonumber \\
&=&(-1)^{n}\frac{\partial (f_{n+1},\;\dots,\;f_{2n},\;f_{1},\;\dots ,\;f_{n})%
}{\partial (q^{1},\;\dots ,\,q^{n},\;p_{1},\dots ,\;p_{n})}\; .  \nonumber
\end{eqnarray}
By the antisymmetry properties of determinant, or in view of $V_{L}(%
\bbox{\xi}_{1},\;\dots ,\;\bbox{\xi}_{2n})=(-1)^{n}V_{L}(\bbox{\xi}%
_{n+1},\;\dots ,\;\bbox{\xi}_{2n},\;\bbox{\xi}_{1},\;\dots ,\;\bbox{\xi}%
_{n}) $, the above relation can be written as 
\begin{equation}
V_{L}(\bbox{\xi}_{f_{1}},\;\dots ,\;\bbox{\xi}_{f_{2n}})= \frac{\partial
(f_{1},\;\dots ,\;f_{2n})}{\partial (q^{1},\;\dots ,\,q^{n},\;p_{1},\dots
,\;p_{n})}\; .
\end{equation}

We now evaluate $\Omega ^{n}$ on the same set of Hamiltonian vector fields
as follows 
\begin{eqnarray}
\Omega ^{n}(\bbox{\xi}_{f_{1}},\;\dots ,\;\bbox{\xi}_{f_{2n}})
&=&\sum_{S_{2n}}(-1)^{\varepsilon } \Omega (\bbox{\xi}_{f_{i_{1}}},\bbox{\xi}%
_{f_{i_{2}}})\cdots \Omega (\bbox{\xi}_{f_{k_{1}}},\bbox{\xi}_{f_{k_{2}}}) 
\nonumber \\
&=&\sum_{S_{2n}}(-1)^{\epsilon }\{f_{i_{1}},f_{i_{2}}\}\;\cdots
\;\{f_{k_{1}},f_{k_{2}}\},
\end{eqnarray}
where $i_{1}<i_{2};\;\dots \;;k_{1}<k_{2}$ such that $(i_{1},i_{2},\;\dots,%
\;k_{1},\;k_{2})$ is a permutation of $(1,\;2,\;\dots ,\;2n)$, and we have
used Eq. (20a). \footnote{%
When we were about to submit this paper we came across to very recent study 
\cite{Zachos2} which derive Eq. (23) (with Jacobian at the left hand side)
without referring to the above symplectic techniques. They successfully
use it in Nambu formulation of a class of systems whose symmetry algebras
close into some simple Lie algebras (see the next section) and then
they develop, in analogy to the classical case, interesting quantum
versions of NB.} Let
us make the identifications 
\begin{equation}
f_{1}=f,\;f_{2}=H\;,\;f_{2+i}=H_{i}\;,\;f_{n+1+i}=A_{i};\;i=1,\;2,\dots
,\;n-1 \;,
\end{equation}
where $H,\;H_{i},\;A_{i}\in SI_{H}(2n-1)$ and $f\in {\cal A}$ is an
arbitrary function. In that case, at the right hand side of Eq. (23) only
the terms in which $H$ is paired with $f$ and each $H_{i}$ is paired with
one of $A_{j}$ give non-zero contributions. All other possible pairing are
zero by very definition of maximal superintegrability. With this in mind let
us consider a fixed partition 
\begin{eqnarray}
\{f_{i_{1}},f_{i_{2}}\}\{f_{j_{1}},f_{j_{2}}\}\;\dots
\;\{f_{k_{1}},f_{k_{2}}\}.  \nonumber
\end{eqnarray}
There are $n$ different places for $\{f_{i_{1}},f_{i_{2}}\}$, and for a
fixed place of $\{f_{i_{1}},f_{i_{2}}\}$ there are $n-1$ possible places for 
$\{f_{j_{1}},f_{j_{2}}\}$, and $n-2$ places for the next pair (provided that
the places of the first two pairs are fixed), and so on. In all this cases
the signs of permutations are the same since each is obtained from the
initial one by interchanging two fixed pairs of indices. Hence, in the right
hand side of (23) there are $n!$ identical copies of each non-zero term.
Thus, (23) can be written, in view of identifications given by (24) , as 
\begin{equation}
\Omega ^{n}(\bbox{\xi}_{f_{1}},\;\dots ,\;\bbox{\xi}_{f_{2n}}) =n!
K(n)\{f,H\}\;,
\end{equation}
where $K(n)$ represents all different partitions which are non-zero 
\begin{equation}
K(n)=N\epsilon ^{i_{1}\;\dots \;i_{n-1}} \{H_{1},A_{i_{1}}\}\;\dots
\;\{H_{n-1},A_{i_{n-1}}\}.
\end{equation}
Since $\epsilon ^{1\;\dots \;n-1}=1$, the factor $N$ is found to be $%
(-1)^{(n-1)(n-2)/2}$ by computing the parity of permutation 
\[
\left( 
\begin{array}{cccccccccc}
1 & 2 & 3 & 4 &  & 5 & 6 & 7\dots & 2n-1 & 2n \\ 
1 & 2 & 3 & n+2 &  & 4 & n+3 & 5\dots & n+1 & 2n
\end{array}
\right) . 
\]
As a result, by defining $B_{ij}=\{H_{i},A_{j}\}$ and the $(n-1)\times (n-1)$
matrix $B=(B_{ij})$ with determinant 
\begin{eqnarray}
detB=\epsilon ^{i_{1}\;\dots \;i_{n-1}} B_{1i_{1}}\;\dots\;B_{n-1\;i_{n-1}} ,
\nonumber
\end{eqnarray}
Eq. (25)can be written as 
\begin{equation}
\Omega ^{n}(\bbox{\xi}_{f_{1}},\;\dots ,\;\bbox{\xi}%
_{f_{2n}})=(-1)^{(n-1)(n-2)/2}n! detB \{f,H\}.
\end{equation}

It will be convenient to give the explicit calculation of (27) for $%
n=1,\;2,\; 3$. For $n=1$ we immediately get $\Omega (\bbox{\xi}_{f_{1}},\;%
\bbox{\xi}_{f_{2}})=\{f,H\}$, and $detB=1$. In the case of $n=2$ we obtain
directly from (23) 
\begin{eqnarray}
\Omega^{2}(\bbox{\xi}_{f_{1}},\;\dots,\; \bbox{\xi}_{f_{4}})=
(-1)^{\epsilon_{1}} \{f_{1},f_{2}\} \{f_{3},f_{4}\}+ (-1)^{\epsilon_{2}}
\{f_{3},f_{4}\}\{f_{1},f_{2}\}\;,  \nonumber
\end{eqnarray}
where $\epsilon_{1}$ is the parity of identity permutation and $\epsilon_{2}$
is the parity of 
\begin{eqnarray}
\left ( 
\begin{array}{cccc}
1 & 2 & 3 & 4 \\ 
3 & 4 & 1 & 2
\end{array}
\right ) = (13)(24)\;\Rightarrow \; \epsilon_{2}=0=\epsilon_{1}.  \nonumber
\end{eqnarray}
Hence $\Omega^{2}(\bbox{\xi}_{f_{1}},\;\dots,\; \bbox{\xi}_{f_{4}})= 2\{f,H
\} B_{11}$, and $detB= B_{11}=\{H_{1},A_{1}\}$. For $n=3$ we have 
\begin{eqnarray}
\Omega^{3}(\bbox{\xi}_{f_{1}},\;\dots,\; \bbox{\xi}_{f_{6}})&=&
\{f_{1},f_{2}\} \sum_{i=1}^{3} \{ [ (-1)^{\epsilon^{1}_{i}}+
(-1)^{\epsilon^{2}_{i}}] \{f_{3},f_{5}\} \{f_{4},f_{6}\}+  \nonumber \\
& & [ (-1)^{\epsilon^{3}_{i}}+ (-1)^{\epsilon^{4}_{i}}]
\{f_{3},f_{6}\}\{f_{4},f_{5}\} \}\;,
\end{eqnarray}
where $\epsilon_{1}^{j}\; ; j=1,\; , 2,\; 3,\;4 $, are for the following
permutations 
\begin{eqnarray}
\left ( 
\begin{array}{cccccc}
1 & 2 & 3 & 4 & 5 & 6 \\ 
1 & 2 & 3 & 5 & 4 & 6
\end{array}
\right ) &= & (45)\;\Rightarrow \; \epsilon^{1}_{1}=1 \;,  \nonumber \\
\left ( 
\begin{array}{cccccc}
1 & 2 & 3 & 4 & 5 & 6 \\ 
1 & 2 & 4 & 6 & 3 & 5
\end{array}
\right ) &=& (3465)\;\Rightarrow \; \epsilon^{2}_{1}=1 \;,  \nonumber \\
\left ( 
\begin{array}{cccccc}
1 & 2 & 3 & 4 & 5 & 6 \\ 
1 & 2 & 3 & 6 & 4 & 5
\end{array}
\right ) &=& (465)\;\Rightarrow \; \epsilon^{3}_{1}=0 \;,  \nonumber \\
\left ( 
\begin{array}{cccccc}
1 & 2 & 3 & 4 & 5 & 6 \\ 
1 & 2 & 4 & 5 & 3 & 6
\end{array}
\right ) &=& (345)\;\Rightarrow \; \epsilon^{4}_{1}=0 \;.  \nonumber
\end{eqnarray}
For $\epsilon_{2}^{i}$ and $\epsilon_{3}^{i}$, the first pair $(12)$ must be
interchanged, respectively, with the second pair and the last pair in the
above permutations; in both cases we have the same signs as $%
\epsilon_{1}^{i} $. Hence Eq. (28) can be written as 
\begin{equation}
\Omega^{3}(\bbox{\xi}_{f_{1}},\;\dots,\; \bbox{\xi}_{f_{6}})=-3!
\{f,H\}(B_{11}B_{22}- B_{12}B_{21}),  \nonumber
\end{equation}
and we get $detB=B_{11}B_{22}- B_{12}B_{21}$.

In view of Eq. (18) a comparison of Eqs. (22) and (27) yields 
\begin{equation}
\frac{\partial(f,\;H,\; H_{1},\;\dots H_{n-1},\;A_{1},\;\dots,\;A_{n-1})} {%
\partial (q^{1},\;\dots,\, q^{n},\;p_{1},\dots,\;p_{n})} =(-1)^{n+1}
detB\{f,H\}\;.
\end{equation}
In terms of 
\begin{eqnarray}
d^{n}x &=& dx^{1}\wedge \; \cdots \;\wedge dx^{n},  \nonumber \\
d^{n}\hat{x}^{i} &=& dx^{1}\wedge \; \cdots \;\wedge dx^{i-1}\wedge
dx^{i+1}\wedge\; \cdots \wedge dx^{n}\;,  \nonumber
\end{eqnarray}
we obtain for $\Gamma$, by comparing (11) and (30) 
\begin{equation}
\Gamma=detB \sum_{i=1}^{n}(-1)^{i-1} [\partial_{q^{i}}Hd^{n}q\wedge d^{n}%
\hat{p}_{i}- (-1)^{n}\partial_{p_{i}}Hd^{n}\hat{q}^{i}\wedge d^{n}p],
\end{equation}
where the summation is explicitly written to avoid any confusion. Note that $%
^{*}(df\wedge \Gamma)$ is equal to the left hand side of Eq. (30). We
conclude this section by the following statement.

If $H\neq $ constant then the constants of motion $H,\; H_{i},\; A_{i}\in
SI_{H}(2n-1)$ are functionally dependent $(\Gamma =0)$ if and only if $%
detB=0 $. These functions are functionally independent $(\Gamma \neq 0)$
where $detB\neq 0$.

\section{Nambu Formulation of Maximally Superintegrable Systems and Their
Symmetry Algebras}

Since the Jacobian determinant (30) is proportional to $\{f,\;H\}$, the
correct time evolution of $f$ can be expressed by the properly normalized NB
as in Eq. (15) with the normalization coefficient $C=(-1)^{n+1}/detB$.
Indeed, the bracket 
\begin{eqnarray}
\{f,H\}_{NP}^{(0)} &=&^{\ast }(df\wedge \Gamma ^{\prime }),  \nonumber \\
&=&\frac{(-1)^{n+1}}{detB}\frac{\partial (f,\;H,\;H_{1},\;\dots,
\;H_{n-1},\;A_{1},\;\dots,\;A_{n-1})}{\partial (q^{1},\;\dots
,\;q^{n},\;p_{1},\;\dots ,\;p_{n})},
\end{eqnarray}
written in terms of ($2n-1$)-form 
\begin{equation}
\Gamma ^{\prime }=\frac{(-1)^{n+1}}{detB}\Gamma
\end{equation}
produces the correct Hamiltonian time evolution 
\begin{equation}
\frac{df}{dt}=\{f,H\}_{NP}^{(0)}=\{f,H\}\;,
\end{equation}
where $f$ is an arbitrary function. Therefore, every $nD$ maximal
superintegrable time-independent Hamiltonian system defined by (1) admits
equivalent Nambu formulation.

It must be emphasized that as we have proved in Sec. III the normalized NB
defined, for two arbitrary functions $h,f\in {\cal A}$, by 
\begin{equation}
\{f,h\}_{NB}^{(0)}=^{\ast }(df\wedge dh\wedge \Gamma _{H})\;,
\end{equation}
obeys the Jacobi identity. Here the $(2n-2)$-form $\Gamma _{H}$ is defined
as 
\begin{equation}
\Gamma _{H}=\frac{(-1)^{n+1}}{detB}dH_{1}\wedge \cdots \wedge dH_{n-1}\wedge
dA_{1}\wedge \cdots \wedge dA_{n-1}.
\end{equation}

The above discussion singles out an important special case in which the
symmetry algebra of a $SI_{H}(2n-1)$ system is such that $detB$ is
everywhere a non-zero constant. In such a case the constants of motion are
globally functional independent and the Nambu formulation is possible
without any (nontrivial) normalization coefficient. For this reason the rest
of this section is devoted to a discussion of some general points of
symmetry algebras of superintegrable systems.

Besides the vanishing PBs given by (1) the defining relations of the
symmetry algebra of a $SI_{H}(2n-1)$ system contain the following 
\begin{equation}
\{H_{i},A_{j}\}=B_{ij},\qquad \{A_{i},A_{j}\}=C_{ij}.
\end{equation}
By the Jacobi identity $B_{ij}$ and $C_{ij}$ are constants of motion and
each is functionally dependent to the constants of motion specified by the
set $SI_{H}(2n-1)$. Hence, each of them can be expressed as $%
X=X(H,\;H_{i},\;A_{j})$. By making use of the identity 
\begin{equation}
\{R,X\}=\{R,H\}\partial _{H}X+\sum_{i=1}^{n-1}\left( \{R,H_{i}\}\partial
_{H_{i}}X+\{R,A_{i}\}\partial _{A_{i}}X\right)
\end{equation}
we obtain 
\begin{mathletters}
\begin{eqnarray}
\{H_{i},X\} &=&\sum_{j=1}^{n-1}B_{ij}\partial _{A_{j}}X, \\
\{A_{i},X\} &=&\sum_{j=1}^{n-1}\left( -B_{ji}\partial
_{H_{j}}X+C_{ij}\partial _{A_{j}}X\right) ,
\end{eqnarray}
Obviously, the symmetry algebra of $SI_{H}(2n-1)$ system is a Lie algebra if
and only if each of $B_{ij}$ and $C_{ij}$ is at most first order in $%
H,\;H_{i}$ and $A_{j}$. If each of $B_{ij}$ and $C_{ij}$ is a polynomial of
degree at most $k>1$, then the symmetry algebra is called to be a polynomial
Poisson algebra of degree $k$. There may also be cases in which the right
hand sides of Eqs. (39) are polynomials but some of $B_{ij}$ and $C_{ij}$
are not. In such cases, by including the non-polynomial ones into the set of
symmetry algebra we again obtain polynomial algebras. Although these
possible cases are by no means exhaustive, especially for low values of $n$
they are likely to occur as interesting structures \cite
{Zhedanov,Vinet,Kalnins,Daskaloyannis}.

\section{ Multi-Hamiltonian Structures of Maximally Superintegrable Systems}

To ease the calculation of this section we shall use the notion $%
H_{k};\;k=0,1,\dots ,2n-2$ such that $H_{0}=H$ and $H_{n+i-1}=A_{i},\;i=1,\;%
\dots ,\;n-1$. Then, in terms of $(2n-2)$-forms 
\end{mathletters}
\begin{equation}
\Gamma _{H_{k}}=\frac{(-1)^{n+k+1}}{detB}dH_{0}\wedge dH_{1}\wedge \cdots
\wedge dH_{k-1}\wedge \hat{dH}_{k}\wedge dH_{k+1}\wedge \cdots \wedge
dH_{2n-2},
\end{equation}
where a hat over a quantity indicates that it should be omitted, we can
define $2n-1$ different normalized NBs as follows 
\begin{equation}
\{f,h\}_{NP}^{(k)}=^{\ast }(df\wedge dh\wedge \Gamma _{H_{k}}),\quad
k=0,\;1,\;\dots ,\;2n-2\;.
\end{equation}
Each of these brackets gives the original time evolution provided that we
choose the new Hamiltonian function to be $H_{k}$: 
\begin{equation}
\frac{df}{dt}=\{f,H_{k}\}_{NP}^{(k)}=\{f,H\}\;.
\end{equation}
In such a case, the system given by (1) has the so-called multi-Hamiltonian
structures property:$\;$ it can equally well be described by any one of the $%
2n-1$ pairs $(H_{k},\;\{,\}_{NP}^{(k)})$.

We shall now prove that the above defined brackets are pairwise compatible,
that is, $\{,\}_{NP}^{(k_{1}k_{2})}=a\{,\}_{NP}^{(k_{1})}+b\{,%
\}_{NP}^{(k_{2})}$ satisfies the Jacobi identity for all $a,b\in {\bf R}$,
independently from the form of $detB$. Evidently, there is no loss of
generality in taking $k_{1}=0,k_{2}=1$ and $a=1=b$. Then let us consider 
\begin{eqnarray}
\{f,\{g,h\}_{NP}^{(01)}\}_{NP}^{(01)}+ cp &=&
[\{f,\{g,h\}_{NP}^{(0)}\}_{NP}^{(1)}+cp]+[\{f,\{g,h\}_{NP}^{(1)}%
\}_{NP}^{(0)}+cp],  \nonumber \\
&=& \{^{*}[df\wedge d^{*}(dg\wedge dh\wedge \Gamma_{H_{0}})\wedge
\Gamma_{H_{1}}]+cp \}+ \\
& &\{^{*}[df\wedge d^{*}(dg\wedge dh\wedge \Gamma_{H_{1}})\wedge
\Gamma_{H_{0}}]+cp \},  \nonumber
\end{eqnarray}
for three functions $f,g$ and $h$. In writing Eq. (43) we have made use of
the fact that $\{,\}_{NP}^{(0)}$ and $\{,\}_{NP}^{(1)}$ satisfy the Jacobi
identity separately. The definition of $\{,\}_{NP}^{(k)}$ requires $2n-1$
independent functions but (43) involves $2n+2$ functions. Therefore, in the
most general case any two of $(f,g,h)$, say $f$ and $g$, must depend on the
other $2n$ independent functions. That is, we can take 
\begin{eqnarray}
df = a_{1}dh+\sum_{k=0}^{2n-2}b_{k}dH_{k},\qquad dg =
a_{2}dh+\sum_{k=0}^{2n-2}c_{k}dH_{k},  \nonumber
\end{eqnarray}
where $a_{1},a_{2},b_{k}, c_{k}$ are arbitrary constants. When these are
substituted in (43), with special care paid to the signs, the right hand
side vanishes.

To point out another important property of superintegrable systems we shall
use local canonical coordinates $x^{\gamma};\;\gamma=1,\dots,2n$ such that $%
x^{j}=q^{j},\;x^{n+j}=p_{j}$ for $j=1,\;\dots ,\;n$. Then, in terms of the
so-called Poisson tensor components 
\begin{eqnarray}
\Lambda _{(k)}^{\alpha \beta } =\frac{(-1)^{n+k+1}}{detB}\varepsilon
^{\alpha \beta \gamma _{0}\;\dots \;\hat{\gamma_{k}}\dots\;\gamma _{2n-2}} 
\frac{\partial H_{0}}{\partial x^{\gamma _{0}}} \frac{\partial H_{1}}{%
\partial x^{\gamma _{1}}} \;\cdots \; \frac{\partial H_{k-1}}{\partial
x^{\gamma _{k-1}}} \hat{\frac{\partial H_{k}}{\partial x^{\gamma _{k}}}} 
\frac{\partial H_{k+1}}{\partial x^{\gamma _{k+1}}} \;\cdots \; \frac{%
\partial H_{2n-2}}{\partial x^{\gamma _{2n-2}}}\;,
\end{eqnarray}
we rewrite Eq. (41) as 
\begin{equation}
\{f,h\}_{NP}^{(k)}=\Lambda _{(k)}^{\alpha \beta }\frac{\partial f}{\partial
x^{\alpha }}\frac{\partial h}{\partial x^{\beta }}\;,\qquad
k=0,\;1,\;\dots,\;2n-2 .
\end{equation}
In Eqs. (44-45) and below summations over repeated Greek letters range from $%
1$ to $2n$. It is not hard to verify that in terms of Poisson tensor the
Jacobi identity means that 
\begin{equation}
\Lambda _{(k)}^{\eta \gamma }\frac{\partial }{\partial x^{\gamma }}\Lambda
_{(k)}^{\alpha \beta }+cp=0\;,
\end{equation}
where $cp$ indicates the cyclic sum with respect to superscripts $\eta
,\;\alpha $ and $\beta $. We shall now prove that all of these Poisson
tensors are singular, that is, $\Lambda _{(k)}$ being the $2n\times 2n$
matrix with elements $\Lambda _{(k)}^{\alpha \beta }$ we have 
\begin{equation}
det(\Lambda _{(k)})=\varepsilon _{\alpha _{1};\dots \;\alpha _{2n}}
\Lambda_{(k)}^{1\alpha _{1}}\;\dots \Lambda _{(k)}^{2n\alpha _{2n}}=0.
\end{equation}

The easiest way to prove Eq. (47) may be as follows. Let $\{{\bf e}_{\alpha
}:\;\alpha =1,\;\dots,\;2n\}$ be a basis of vector space ${\bf R}^{2n}$ and
let us define $2n$ vectors ${\bf v}^{(\alpha )}=\Lambda _{(0)}^{\alpha \beta
}{\bf e}_{\beta }$, where , without any loss of generality, we have taken $%
k=0$. Then for $k^{\prime} =1,\;\dots ,\;2n-2$ we have 
\begin{eqnarray}
{\bf v}^{(\alpha )}\cdot \bbox{\nabla}H_{k^{\prime} } &=&\Lambda
_{(0)}^{\alpha \beta }\frac{\partial H_{k^{\prime} }}{\partial x^{\beta }} 
\nonumber \\
&=&\varepsilon ^{\alpha \beta \gamma _{1}\;\dots \;\gamma _{2n-2}}\frac{%
\partial H_{1}}{\partial x^{\gamma _{1}}}\;\cdots \;\frac{\partial H_{2n-2}}{%
\partial x^{\gamma _{2n-2}}}\frac{\partial H_{k^{\prime} }}{\partial
x^{\beta }}=0\;,  \nonumber
\end{eqnarray}
because of the contraction of two symmetric and two antisymmetric indices.
This proves that each of ${\bf v}^{(\alpha )}$ is perpendicular to the set
of $2n-2$ linearly independent vectors $S_{H}=\{\bbox{\nabla}H_{k^{\prime}
}:\; k^{\prime} =1,\;\dots ,\;2n-2\}$. Hence the $2n$ vectors ${\bf v}%
^{(\alpha )}$ are linearly dependent and therefore the matrix of their
components, which is the matrix $\Lambda _{(0)}$, is singular. In fact the
rank of this matrix is two since this is the dimension of the orthogonal
complement of the set $S_{H}$.

Evidently, all of the Poisson tensors defined above are pairwise compatible,
but as they are singular they do not lead to any symplectic structure.

\section{Applications}

\subsection{Calogero-Moser system}

The Calogero-Moser system is one of the four $nD$ systems which are known to
be maximally superintegrable for any finite integer $n$ \cite
{Gonera,Evans,Wol}. The other three systems are the Kepler-Coulomb problem,
harmonic oscillator with rational frequency ratios, and Winternitz system.
Here we shall consider $n=2$ (two particles) case of the Calogero-Moser
system described by 
\begin{equation}
H_{CM}=\frac{1}{2m}{\bf p}^{2}+\frac{g^{2}}{2(q^{1}-q^{2})^{2}}\; ,
\end{equation}
where $g$ is a constant. Constants of motion for $H_{CM}$ can be written as 
\begin{eqnarray}
H_{1} &=&2(q^{1}+q^{2})H_{CM}-{\bf q}\cdot {\bf p}\frac{p_{1}+p_{2}}{m}, 
\nonumber \\
A_{1} &=&\frac{p_{1}+p_{2}}{m}, \\
A_{1}^{\prime } &=&\frac{1}{2m^{2}}(p_{1}-p_{2})^{2}+{\frac{g^{2}}{%
m(q^{1}-q^{2})^{2}}}.  \nonumber
\end{eqnarray}
They obey the following relations 
\begin{eqnarray}
\{H_{CM},H_{1} \} &=& \{H_{CM}, A_{1} \}=\{H_{CM},A_{1}^{\prime}\} =0, 
\nonumber \\
\{A_{1},A_{1}^{\prime }\} &=&0,\quad \{H_{1},A_{1} \} =2A_{1}^{\prime
},\quad \{H_{1},A_{1}^{\prime }\}=-2A_{1}A_{1}^{\prime }\;.
\end{eqnarray}

The relation $\{A_{1},A_{1}^{\prime }\}=0$ implies (and is implied by) the
fact that $H_{CM},\;A_{1}$ and $A_{1}^{\prime }$ are functionally dependent.
From Eqs. (48) and (49) one can easily identify this dependence as $%
A_{1}^{\prime }=(4H_{CM}-mA_{1}^{2})/2m$. Thus, as a functional independent
set we can take $(H_{CM},\;H_{1},\;A_{1})$, or $(H_{CM},\;H_{1},\;A_{1}^{%
\prime })$. In the former case the symmetry algebra is spanned by $%
(H_{CM},\;H_{1},\;A_{1})$ and since 
\begin{equation}
B_{11}=\{H_{1},A_{1} \}= -A_{1}^{2}+\frac{4}{m}H_{CM},
\end{equation}
it is a quadratic Poisson algebra. In the latter case the symmetry algebra
is spanned by $(H_{CM},\;H_{1},\;A_{1}^{\prime },\;B_{11}^{\prime})$, where $%
B_{11}^{\prime}=\{H_{1},A_{1}^{\prime }\}$ such that 
\begin{equation}
B_{11}^{\prime 2}=-8A_{1}^{\prime 3}+\frac{16}{m}A_{1}^{\prime 2}H_{CM}.
\end{equation}
From Eqs. (39a) and (39b) it is found that 
\begin{mathletters}
\begin{eqnarray}
\{H_{1},B_{11}^{\prime} \} &=& \frac{1}{2}\frac{\partial B_{11}^{\prime 2} }{%
\partial A_{1}^{\prime}} =-12A_{1}^{\prime 2}+\frac{16}{m}%
A_{1}^{\prime}H_{CM}, \\
\{A_{1}^{\prime},B_{11}^{\prime} \} &=& -\frac{1}{2}\frac{\partial
B_{11}^{\prime 2} }{\partial H_{1}}=0.
\end{eqnarray}
Hence we have again a quadratic Poisson algebra.

In both cases the time evolutions can be written in the Nambu and
Hamiltonian mechanics equivalently as 
\end{mathletters}
\begin{mathletters}
\begin{eqnarray}
\frac{df}{dt} &=&-\frac{1}{B_{11}}\{f,H_{CM},H_{1},A_{1} \} \\
&=&-\frac{1}{B_{11}^{\prime }}\{f,H_{CM},H_{1},A_{1}^{\prime
}\}=\{f,H_{CM}\}\;.
\end{eqnarray}
Both of the brackets 
\begin{eqnarray}
\{f,g\}^{(0)}_{NP} &=&-\frac{1}{B_{11}}\{f,g,H_{1},A_{1} \}\;,  \nonumber \\
\{f,g\}^{\prime (0)}_{NP} &=&-\frac{1}{B_{11}^{\prime }}\{f,g,H_{1},A_{1}^{%
\prime }\}\;,  \nonumber
\end{eqnarray}
defined for two arbitrary functions $f,\;g$ satisfy the Jacobi identity.
This is also the case for the normalized NBs corresponding to the
Hamiltonian functions $H_{1}$ 
\begin{eqnarray}
\{f,g\}^{(1)}_{NP} &=&\frac{1}{B_{11}}\{f,g,H_{CM},A_{1} \}\;,  \nonumber \\
\{f,g\}^{\prime (1)}_{NP} &=&\frac{1}{B_{11}^{\prime }}\{f,g,H_{CM},A_{1}^{%
\prime }\}\;,  \nonumber
\end{eqnarray}
and to $A_{1}$ and to $A_{1}^{\prime}$ 
\begin{eqnarray}
\{f,g\}^{(2)}_{NP} &=&-\frac{1}{B_{11}}\{f,g,H_{CM},H_{1} \}\;,  \nonumber \\
\{f,g\}^{\prime (2)}_{NP} &=&-\frac{1}{B_{11}^{\prime }}\{f,g,H_{CM},H_{1}^{%
\prime }\}\;.  \nonumber
\end{eqnarray}

Finally in this subsection we should note that by redefinition of the
constants of motion we can make the normalization coefficients trivial at
the expense of restricting their domains of definition. As an example we
consider the constant of motion 
\end{mathletters}
\[
A^{\prime \prime }=\frac{1}{2h}ln\frac{A_{1}-h}{A_{1}+h},\quad h=2\sqrt{%
\frac{H_{CM}}{m}}, 
\]
which satisfies $\{H_{1},A_{1}^{\prime \prime }\}=\{H_{1},A_{1}\}\partial
_{A_{1}}A^{\prime \prime }=-1$. Then, we can rewrite Eq. (54a) as 
\begin{equation}
\frac{df}{dt}=\{f,H_{CM},H_{1},A^{\prime \prime }\}=\{f,H_{CM}\}.
\end{equation}
Two more alternative brackets can be defined as 
\begin{eqnarray}
\{f,g\}_{NP}^{\prime \prime (1)} &=&-\{f,g,H_{CM},A_{1}^{\prime \prime }\}\;,
\nonumber \\
\{f,g\}_{NP}^{\prime \prime (2)} &=&\{f,g,H_{CM},H_{1}\}\;.  \nonumber
\end{eqnarray}
Like others, all these brackets obey the Jacobi identity are pairwise
compatible and each is degenerate. Explicit forms of corresponding Poisson
tensors can also be found as will be done for the next application.

\subsection{Landau problem}

We now consider the well-known Landau Hamiltonian $H_{L} $ and first
establish its superintegrability and Nambu formulation. Explicit expressions
of three different Hamiltonian structures of this problem will be presented
in the next subsection.

For a particle of charge $q>0$ and mass $m$ moving on the $q^{1}q^{2}$-plane
under the influence of the perpendicular static and uniform magnetic field $%
{\cal B}=\partial _{q^{1}}a_{2}-\partial _{q^{2}}a_{1}$ , $H_{L}$ is (in the
Gaussian units) 
\begin{equation}
H_{L}=\frac{1}{2m}({\bf p}-\frac{q}{c}{\bf a})^{2}=\frac{1}{2}%
m(v_{1}^{2}+v_{2}^{2}),
\end{equation}
where $c$ is the speed of light, ${\bf a}=(a_{1},\;a_{2})$ is the vector
potential and ${\bf v}=({\bf p}-\frac{q}{c}{\bf a})/m$ is the velocity
vector \cite{Landau}. Components of ${\bf v}$ obey $\{v_{1},v_{2}\}=q{\cal B}%
/m^{2}c$ for any ${\cal B}$.

When ${\cal B}$ is constant the most general form of the vector potential is 
${\bf a}=({\cal B}/2)(-q^{2},\;q^{1})+\bbox{\nabla}_{q}\chi $ , where $\chi
\equiv \chi ({\bf q})$ is an arbitrary gauge function. In such a case we
have, in any gauge $\chi $, two constants of motion 
\begin{equation}
H_{1}=m(v_{2}+\omega q^{1}),\qquad A_{1}=-m(v_{1}-\omega q^{2}),
\end{equation}
where $\omega =q{\cal B}/mc$ is the cyclotron frequency. $%
(H_{1},\;A_{1})/m\omega $ correspond to coordinates of the cyclotron centre
and they satisfy the gauge-independent relations $\{v_{j},H_{1}\}=0=%
\{v_{j},A_{1}\},\;j=1,2$, and 
\begin{equation}
\{H_{L},H_{1}\}=0=\{H_{L},A_{1}\},\quad \{H_{1},A_{1}\}=-m\omega .
\end{equation}
These relations explicitly show that $H_{L},H_{1}$ and $A_{1}$ are
functional independent constants of motion and they close into a Lie algebra
structure which can be identified as the centrally extended Heisenberg-Weyl
algebra. This completes the superintegrability of the Landau problem.

Let us now consider the 3-form $\Gamma =(m\omega )^{-1}dH_{L}\wedge
dH_{1}\wedge dA_{1}$, which can be written as 
\begin{equation}
\Gamma =(v_{1}dq^{2}-v_{2}dq^{1})\wedge \lbrack dp_{1}\wedge dp_{2}-\frac{q}{%
c}(dp_{1}\wedge da_{2}-dp_{2}\wedge da_{1})]+m\omega {\bf v}\cdot d{\bf p}%
\wedge dq^{1}\wedge dq^{2}.
\end{equation}
Making use of this we immediately have 
\begin{eqnarray}
{}^{\star }(d{\bf q}\wedge \Gamma ) &=&{\bf v}\;,  \nonumber \\
{}^{\star }(d{\bf p}\wedge \Gamma ) &=&\frac{q}{c} (v_{1}\bbox{\nabla }%
a_{1}+v_{2}\bbox{\nabla}a_{2})\;.  \nonumber
\end{eqnarray}
It is straightforward to check that the right hand sides of these equations
are the right hand sides of the canonical Hamiltonian equations for $H_{L}$.
Therefore, we can write them collectively as 
\begin{equation}
\frac{d}{dt}{\bf u}={}^{\star }(d{\bf u}\wedge \Gamma ),
\end{equation}
where ${\bf u}=(q^{1},\;q^{2},\;p_{1},\;p_{2})$. One may also write Eq. (60)
for any function $f$.

\subsection{Three Hamiltonian Structures of Landau Problem}

Since $B_{11}=\{H_{1},A_{1}\}=-m\omega$ we can define the following three
different Poisson tensors 
\begin{eqnarray}
\Lambda^{\alpha \beta}_{(0)} & = & \frac{1}{m\omega} \varepsilon ^{\alpha
\beta \gamma _{1}\gamma _{2}}\frac{\partial H_{1}}{\partial x^{\gamma _{1}}}%
\frac{\partial A_{1}}{\partial x^{\gamma _{2}}}\;,  \nonumber \\
\Lambda^{\alpha \beta}_{(1)} & = & -\frac{1}{m\omega} \varepsilon ^{\alpha
\beta \gamma _{1}\gamma _{2}}\frac{\partial H_{L}}{\partial x^{\gamma _{1}}}%
\frac{\partial A_{1}}{\partial x^{\gamma _{2}}}\;, \\
\Lambda^{\alpha \beta}_{(2)} & = & \frac{1}{m\omega} \varepsilon ^{\alpha
\beta \gamma _{1}\gamma _{2}}\frac{\partial H_{L}}{\partial x^{\gamma _{1}}}%
\frac{\partial H_{1}}{\partial x^{\gamma _{2}}}\;,  \nonumber
\end{eqnarray}
for the Landau problem. These are all computed and in terms of $2\times 2$
matrices 
\begin{eqnarray}
Y=\left ( 
\begin{array}{cc}
0 & 1 \\ 
-1 & 0
\end{array}
\right ),\quad Z=\left ( 
\begin{array}{cc}
\partial_{2} a_{1} & \partial_{2} a_{2} \\ 
-\partial_{1} a_{1} & -\partial_{1} a_{2}
\end{array}
\right ),\quad S_{1}=\left ( 
\begin{array}{cc}
0 & v_{1} \\ 
0 & v_{2}
\end{array}
\right ),\quad S_{2}=\left ( 
\begin{array}{cc}
-v_{1} & 0 \\ 
-v_{2} & 0
\end{array}
\right ),
\end{eqnarray}
and 
\begin{eqnarray}
l=det Z\;, \quad l_{1}= v_{1}\partial_{1} a_{1} +v_{2}\partial_{1} a_{2}\;,
\quad l_{2}= v_{1}\partial_{2} a_{1} +v_{2}\partial_{2} a_{2}\;,
\end{eqnarray}
they can be expressed by the following $4\times 4$ matrices 
\begin{eqnarray}
\Lambda_{(0)} &=&J_{0}+\frac{1}{{\cal B} }\left ( 
\begin{array}{cc}
\frac{c}{q}Y & Z \\ 
-\tilde{Z} & \frac{q}{c}lY
\end{array}
\right )\;,  \nonumber \\
\Lambda_{(1)} &=&-v_{2}\Lambda_{(0)}+\left ( 
\begin{array}{cc}
{\bf 0} & S_{1} \\ 
-\tilde{S_{1}} & \frac{q}{c}l_{1}Y
\end{array}
\right )\;, \\
\Lambda_{(2)} &=& v_{1}\Lambda_{(0)}+\left ( 
\begin{array}{cc}
{\bf 0} & S_{2} \\ 
-\tilde{S_{2}} & \frac{q}{c}l_{2}Y
\end{array}
\right )\;.  \nonumber
\end{eqnarray}
Here $\tilde{Z} $ denotes the transposition of the matrix $Z$ and in terms
of $2\times 2$ zero matrix ${\bf 0}$ and unit matrix ${\bf 1}$ 
\begin{equation}
J_{0}=\left ( 
\begin{array}{cc}
{\bf 0} & {\bf 1} \\ 
-{\bf 1} & {\bf 0}
\end{array}
\right ),
\end{equation}
stands for the standard $4\times 4$ symplectic matrix.

We first should note that these Hamiltonian structures are valid in any
gauge $\chi$. Then, as applications of general statements proved in the main
text one can explicitly verify the following.\newline
(i) Each of $\Lambda _{(k)}$ provides the correct Hamiltonian equations for
the Landau problem which can be cast in the following matrix forms 
\begin{equation}
\frac{d}{dt}\tilde{{\bf u}}=\Lambda _{(k)}{\bf \nabla }H_{k}\;;\quad
k=0,\;1,\;2\;,
\end{equation}
where $H_{0}=H_{L},\;H_{2}=A_{1}$ and ${\bf \nabla }H_{k}$ is the column
matrix of gradient $H_{k}$.\newline
(ii) $det\Lambda _{(k)}=0$ and in fact each of $\Lambda _{(k)}$ has rank two.%
\newline
(iii) All $\Lambda _{(k)}$ satisfy the Jacobi identity (46) and they are
pair-wise compatible.

\subsection{Smorodinsky-Winternitz potentials}

We now consider a set of four potentials first found by Winternitz and
co-workers \cite{Winternitz}. Relaying on the assumptions; (i) Hamiltonians
are of potential form, (ii) integrals of motion are at most quadratic in
momenta, they found the following four potentials 
\begin{mathletters}
\begin{eqnarray}
V^{(1)}&=& \frac{1}{2}kr^{2} +\frac{1}{2}\left( \frac{\alpha _{1}}{q_1^{2}}+ 
\frac{\beta_{1}}{q_2^{2}}\right), \\
V^{(2)}&=&\omega(4q_1^{2}+q_2^{2})+\alpha_{2}q_1+\frac{\beta _{2}}{q_2^{2}}%
\;, \\
V^{(3)}&=&\frac{1}{2r}\left(\kappa+ \frac{\alpha_{3}}{r+q_1}+ \frac{\beta_{3}%
}{r-q_1}\right), \\
V^{(4)}&=&\frac{1}{2r}(\sigma+ \alpha_{4}\sqrt{r+q_1}+\beta_{4}\sqrt{r-q_1}),
\end{eqnarray}
where $k, \omega,\kappa,\sigma$ and $\alpha_{j}, \beta_{j}$ are some real
constants and $r^2 = q_1^{2}+q_2^{2}$. All these potentials accept
separation of variables in at least two coordinate systems and each has, in
addition to Hamiltonian function $H^{(j)}=({p}^{2}/2m)+V^{(j)} $, two
constants of motion $H_{1}^{(j)}, A_{1}^{(j)} ; j=1, 2, 3, 4$. That is, all
the Smorodinsky-Winternitz systems are superintegrable and they contain
two-dimensional harmonic oscillator and Kepler-Coulomb problem as special
cases. $D$-dimensional version of (67a) for $D\geq 2$ is known as the
Winternitz system \cite{Evans2}.

Explicit expressions of constants of motion and their nonvanishing PBs are
given altogether in the Table I. In addition to that given in third column
of the table one must add the following relations 
\end{mathletters}
\begin{equation}
\{H_{1}^{(j)},\;H^{(j)}\}=\{A_{1}^{(j)},\;H^{(j)}\}=\{B_{11}^{(j)},\;H^{(j)}%
\}=0,
\end{equation}
to the defining relations of symmetry algebras. The constants of motion $%
B_{11}^{(j)}=\{H_{1}^{(j)},\;A_{1}^{(j)}\}$ are found to be 
\begin{eqnarray}
B_{11}^{(1)} &=&-\frac{4}{m}\left[ L\left( \frac{p_{1}p_{2}}{m}%
+kq_{1}q_{2}\right) -\alpha _{1}\frac{q_{2}p_{2}}{{q_{1}}^{2}}+\beta _{1}%
\frac{q_{1}p_{1}}{{q_{2}}^{2}}\right] ,  \nonumber \\
B_{11}^{(2)} &=&-\frac{2}{m}q_{2}p_{2}\left( 8\omega q_{1}+\alpha
_{2}\right) -\frac{2p_{1}}{m}\left( \frac{p_{2}^{2}}{m}-2\omega q_{2}^{2}+%
\frac{2\beta _{2}}{q_{2}^{2}}\right) , \\
B_{11}^{(3)} &=&-\frac{2p_{1}L^{2}}{m^{2}}+\frac{2q_{2}L}{mr}\left( \frac{%
\kappa }{2}+\frac{\alpha _{3}}{r+q_{1}}+\frac{\beta _{3}}{r-q_{1}}\right) +%
\frac{q_{2}^{2}}{mr}{\bf q}\cdot {\bf p}\left[ \frac{\alpha _{3}}{%
(r+q_{1})^{2}}-\frac{\beta _{3}}{(r-q_{1})^{2}}\right] ,  \nonumber
\end{eqnarray}
where $B_{11}^{(4)}$ is not written due to its length and we have defined 
\begin{equation}
\gamma _{j}^{\pm }=\alpha _{j}\pm \beta _{j},\quad L=p_{2}q_{1}-p_{1}q_{2}.
\end{equation}
Since $B_{11}^{(j)}$s are cubic polynomials in the momenta they can not be
written as polynomials in the constants of motion \cite{Daskaloyannis}. But
their squares can be expressed as follows 
\begin{eqnarray}
B_{11}^{(1)\;2} &=&-\frac{16}{m}\left[ A_{1}^{(1)}(H_{1}^{(1)%
\;2}-2H_{1}^{(1)}H^{(1)}+kA_{1}^{(1)}-2\gamma _{1}^{+}k)+4\alpha
_{1}H^{(1)\;2}\right]  \nonumber \\
&&\frac{16}{m}\gamma _{1}^{-}(2H_{1}^{(1)}H^{(1)}-\gamma _{1}^{-}k), 
\nonumber \\
B_{11}^{(2)\;2} &=&\frac{8}{m}\left[ 4H_{1}^{(2)}(H_{1}^{(2)}-H^{(2)})^{2}+%
\alpha _{2}A_{1}^{(2)}(H_{1}^{(2)}-H^{(2)})-\omega A_{1}^{(2)\;2}-\beta
_{2}(16\omega H_{1}^{(2)}+\alpha _{2}^{2})\right] ,  \nonumber \\
B_{11}^{(3)\;2} &=&-\frac{1}{m}\left[ H_{1}^{(3)}\left(
4A_{1}^{(3)\;2}-8H_{1}^{(3)}H^{(3)}+8\gamma _{3}^{+}H^{(3)}-\kappa
^{2}\right) -2\gamma _{3}^{-}(\kappa A_{1}^{(3)}+\gamma
_{3}^{-}H^{(3)})+\kappa ^{2}\gamma _{3}^{+}\right] ,  \nonumber \\
B_{11}^{(4)\;2} &=&\frac{1}{m}\left[ H_{1}^{(4)}(2H^{(4)}H_{1}^{(4)}-\frac{1%
}{2}\gamma _{4}^{+}\gamma _{4}^{-})+A_{1}^{(4)}(2H^{(4)}A_{1}^{(4)}-\alpha
_{4}\beta _{4})-\frac{\sigma }{2}\left( H^{(4)}+\frac{\alpha _{4}^{2}+\beta
_{4}^{2}}{2}\right) \right] .  \nonumber
\end{eqnarray}
As a result, the symmetry algebras for all the Smorodinsky-Winternitz
potentials are five dimensional quadratic Poisson algebras generated by $%
H^{(j)},\;H_{1}^{(j)},\;A_{1}^{(j)},\;B_{11}^{(j)}$ and $1$. Finally in this
section we will be content with writing the normalized NB of this system as 
\begin{equation}
\frac{df}{dt}=-\frac{1}{B_{11}^{(j)}}{}^{\star }(df\wedge dH^{(j)}\wedge
dH_{1}^{(j)}\wedge dA_{1}^{(j)})=\{f,\;H^{(j)}\}.
\end{equation}
It is also straightforward to write out the alternative Poisson structures
for each member of this class of potentials.

\acknowledgments
We thank \c{S}. Kuru for useful conversations and for her help in preparing
the manuscript of this study. This work was supported in part by the
Scientific and Technical Research Council of Turkey (T\"{U}B\.{I}TAK).

\begin{table}[tbp]
\caption{ Constants of motion and nonvanishing PBs for the symmetry algebras
of the Smorodinsky-Winternitz potentials. The abbreviations $\protect\gamma%
^{\pm}_{j}$ and $L$ are defined by Eq. (70).}
\label{tab1}
\begin{tabular}{|c|l|l|}
& Constants of Motion & Nonvanishing PBs \\ 
\tableline \tableline & $H_{1}^{(1)}=\frac{p_1^{2}}{m}+kq_1^{2}+\frac{%
\alpha_1}{q_1^{2}}$ & $\{ H_{1}^{(1)},B_{11}^{(1)}\}= -\frac{8}{m}%
[H_{1}^{(1)}(H_{1}^{(1)}-2H^{(1)}) $ \\ 
$H^{(1)}$ &  & $\qquad \qquad \qquad+2k(A_{1}^{(1)}-\gamma^{+}_{1})]$ \\ 
& $A_{1}^{(1)}=\frac{L^{2}}{m}+r^{2}\left(\frac{\alpha_1}{q_1^{2}}+ \frac{%
\beta_1}{q_2^{2}}\right) $ & $\{ A_{1}^{(1)},B_{11}^{(1)}\}= \frac{16}{m}%
(H_{1}^{(1)}A_{1}^{(1)}- H^{(1)}A_{1}^{(1)}-\gamma^{-}_{1}H^{(1)})$ \\ 
\tableline & $H_{1}^{(2)}={\frac{1}{2m}}p_1^{2}+4\omega
q_1^{2}+\alpha_{2}q_1 $ & $\{ H_{1}^{(2)},B_{11}^{(2)}{\}}= \frac{4}{m}%
(\alpha_{2}H_{1}^{(2)}-2\omega A_{1}^{(2)}-\alpha_{2}H^{(2)})$ \\ 
$H^{(2)}$ &  &  \\ 
& $A_{1}^{(2)}=\frac{2}{m}Lp_{2}-q_2^{2}(4\omega q_{1}+\alpha_{2})+ {\frac{%
4\beta _{2}q_{1}}{q_2^{2}}} $ & $\{ A_{1}^{(2)},B_{11}^{(2)}{\}} =-\frac{16}{%
m}(3{H_{1}^{(2)}}^{2} -4H^{(2)}H_{1}^{(2)}+{H^{(2)}}^{2}$ \\ 
&  & $\qquad\qquad \qquad+\frac{\alpha_{2}}{4}A_{1}^{(2)} -4\omega\beta_{2})$
\\ 
\tableline & $H_{1}^{(3)}=\frac{1}{m}L^{2}+r\left( {\frac{\alpha_{3}}{{r+q_1}%
}}+ {\frac{\beta_{3}}{{r-q_1}}}\right) $ & $\{ H_{1}^{(3)},B_{11}^{(3)}{\}}=-%
\frac{1}{m}(4H_{1}^{(3)}A_{1}^{(3)}-\kappa \gamma^{-}_{3})$ \\ 
$H^{(3)}$ &  &  \\ 
& $A_{1}^{(3)}=\frac{L}{m}p_{2}-\frac{1}{2r}\left( \alpha_{3}\frac{r-q_1}{%
r+q_1} +\beta_{3}\frac{r+q_1}{r-q_1}+ \kappa q_{1} \right)$ & $\{
A_{1}^{(3)},B_{11}^{(3)}{\}}= \frac{2}{m}[{A_{1}^{(3)}}%
^{2}-2H^{(3)}(2H_{1}^{(3)}- \gamma^{+}_{3})$ \\ 
&  & \qquad \qquad \qquad $- {\frac{1}{4}}\kappa^{2}]$ \\ 
\tableline & $H_{1}^{(4)}=\frac{1}{2r} \left[ \sigma
q_{2}-\alpha_{4}(r-q_{1})\sqrt{r+q_1} \right] $ &  \\ 
& $\qquad +\frac{L}{m}p_{2} +\frac{\beta_{4}}{2r}(r+q_{1})\sqrt{r-q_1} $ & $%
\{ H_{1}^{(4)},B_{11}^{(4)}{\}}=\frac{2}{m}(H^{(4)}A_{1}^{(4)} -{\frac{1}{4}}%
\alpha_{4}\beta_{4})$ \\ 
$H^{(4)}$ &  &  \\ 
& $A_{1}^{(4)}=-\frac{q_{1}}{2r}\left( \alpha_{4}\sqrt{r-q_1}-\beta_{4}\sqrt{%
r+q_1}\right)$ & $\{ A_{1}^{(4)},B_{11}^{(4)}{\}} =-\frac{2}{m}%
(H_{1}^{(4)}H^{(4)} -{\frac{1}{8}}\gamma^{+}_{4}\gamma^{-}_{4})$ \\ 
& $\qquad +\frac{L}{m}p_{1} - \frac{\sigma q_{2}}{2r} $ & 
\end{tabular}
\end{table}

\end{document}